\documentstyle[aps,preprint,prb,epsf]{revtex}
\begin{document}
\draft
\title{On the exciton binding energy in a quantum well}
\author{B. Gerlach\cite{a}, J. W\"usthoff}
\address{Institut f\"ur Physik, Universit\"at Dortmund, D-44221
Dortmund, Germany}
\author{M. O. Dzero, M. A. Smondyrev}
\address{Bogoliubov Laboratory of Theoretical Physics,
Joint Institute for Nuclear Research \\
14980 Dubna, Moscow Region, Russia}
\date{\today}
\maketitle
\begin{abstract}
We consider a model describing the one-dimensional confinement of an
exciton in a symmetrical, rectangular quantum-well structure and derive
upper and lower bounds for the binding energy $E_b$ of the exciton.
Based on these bounds, we study the dependence of $E_b$ on the width of
the confining potential with a higher accuracy than previous reports.
For an infinitely deep potential the binding energy varies as expected
from  $1\,Ry$ at large widths to $4\,Ry$ at small widths. For a finite
potential, but without consideration of a mass mismatch or a dielectric
mismatch, we substantiate earlier results that the binding energy
approaches the value $1\,Ry$ for both small and large widths, having a
characteristic peak for some intermediate size of the slab. Taking the
mismatch into account, this result will in general no longer be true.
For the specific case of a $Ga_{1-x}Al_{x}As/GaAs/Ga_{1-x}Al_{x}As$
quantum-well structure, however, and in contrast to previous findings,
the peak structure is shown to survive.  \end{abstract} \pacs{PACS
numbers: 73.20.Dx + 71.35.-y}

\narrowtext

\section{Introduction and statement of the problem}

The study of electronic and excitonic properties in quantum well
structures has been a subject of great interest since the pioneering
work of Dingle, Wiegmann and Henry \cite{Dingle}. In view of the
enormous amount of literature in this field, any list of references
must be incomplete. We quote the publications
\cite{Burstein,Altarelli,Andrea,Bastard,Mendez,Bajaj,%
Andreani,Winkler,Fomin,Tagakahara,Zimmermann}, which are in part
related to this work, and recommend the references therein. One of the
most appealing features of these systems is the enhancement of
excitonic effects, for instance the increase of the binding energy and
the oscillator strength, which may allow the observation of an exciton
even up to room temperature.

The main reason for the large binding energies and oscillator strengths
is understood to be the quantum confinement of electron and hole in the
growth direction of the heterostructure; in comparison to a
three-dimensional monostructure, the electron-hole correlation is
increased. Moreover, the mass mismatch as well as the dielectric
mismatch in a quantum well may even enhance this effect, as was
primarily pointed out by Keldysh\cite{Keldysh}.

In contrast to the simple qualitative reasonings, a quantitative
theoretical description of the excitonic enhancement is quite
complicated. The reasons are obvious; as translation symmetry is broken
in the growth direction of the heterostructure, the familiar separation
of center-of-mass and relative part of the exciton motion is no longer
possible. If a dielectric mismatch is to be included, the electron-hole
potential is no longer a simple Coulomb potential. If a mass mismatch
exists, the kinetic-energy part of the Hamiltonian is no longer
isotropic. Turning to real substances such as
$Ga_{1-x}Al_{x}As/GaAs/Ga_{1-x}Al_{x}As$, band-structure complications
(e.g. valence-band degeneracy) do occur.  Moreover, the growing process
may induce an interface roughness etc..  As a consequence, the spectra
of excitons in a quantum well are far from being understood on a
quantitative scale and different aspects have  been controversially
discussed in the literature.

In this paper we are concerned with one of the problems which --- to
the best of our knowledge --- has not been solved; i.e.  the position
of the peak of the excitonic binding energy as a function of the width
of the confining potential. Some introductory comments may be
appropriate. For an infinitely deep well, the binding energy is known
to vary monotonically from $1\,Ry$ to $4\,Ry$ if the width $L$ varies
from infinity to zero.  Contrasting this case with that of a finite
well in an otherwise homogeneous material, the binding energy behaves
similarly for sufficiently large values of $L$, but qualitatively
different for small $L$. In the latter case, the wave function spills
increasingly over the interfaces and into the barriers, occupying a
greater three-dimensional volume as tunneling becomes more and more
important. In the ultimate limit of zero width, the binding energy will
again be $1\,Ry$. For a finite well width the confinement will increase
the binding energy $E_b$ above $1\,Ry$.  Thus we are lead to conclude
that $E_b$ should have a maximum for some intermediate well width. In
fact, this behavior was found in a variational treatment due to Greene,
Bajaj, and Phelps \cite{Bajaj}. In Sec.~IV, we complement their results
by lower bounds for the binding energy, which have a relative deviation
of (at most) $0.2$ from the corresponding upper bounds and,
furthermore, a similar shape.

A finite well in an otherwise homogeneous material should probably be
viewed as a rather poor model for a real quantum well. We anticipate,
however, that such a conclusion might be somewhat pessimistic.  The
model is well applicable as starting point for a quantitative
description, but we include mismatches for the masses and the
dielectric constants. What about the binding energy under these
circumstances?  The situation is clear in the limiting cases; for
infinite well width, we start with $1\,Ry$ of the well material,
whereas for zero well width, we arrive at $1\,Ry'$ of the barrier
material. Whether or not a peak appears will be sensitive to the value
of $Ry/Ry'$.  Andreani and Pasquarello \cite{Andreani} performed a
study for $Ga_{1-x}Al_{x}As/GaAs/Ga_{1-x}Al_{x}As$, including the
effects of, firstly, a dielectric mismatch and, secondly, the valence
band degeneracy.  They did not find a peak for $x>0.25$ and a width
$L>30\, \AA$, even when the band degeneracy is (theoretically) switched
off.

The intention of this paper is to critically reexamine this conclusion.
Clearly, our model must also resort to simplifications, as was
indicated above. For each part of the heterostructure, we assume
nondegenerate, isotropic and parabolic bands, but include a mass
mismatch and a dielectric mismatch at the interfaces. The confinement
of electron and hole is mimicked by finite rectangular wells. Our
strategy is to produce upper and lower bounds for the correct binding
energy of the model.  Consequently, we can estimate the error of our
results quantitatively.  Inserting the material parameters of
$Ga_{1-x}Al_{x}As/GaAs/Ga_{1-x}Al_{x}As$, $x$ varying from 0.15 to
0.40, we do find a peak structure. We can thereby disprove the above
assertions.

\section{Formulation of the Model}
\begin{figure}
\phantom{a}\vspace*{-2cm}\hspace*{2cm}
\epsfxsize=18cm \epsfbox{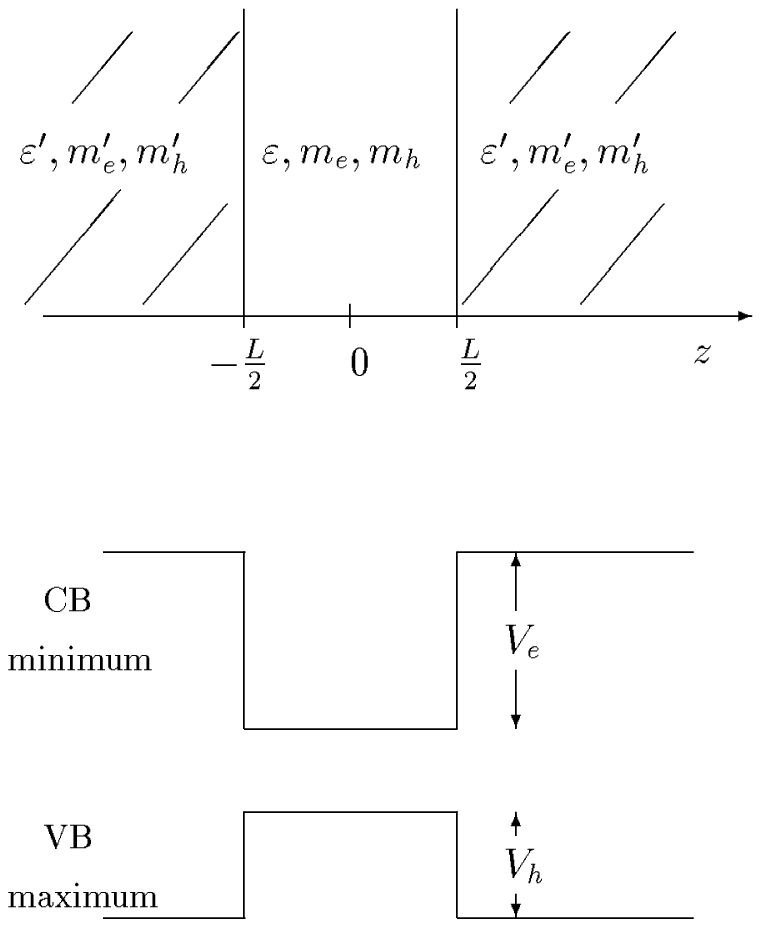}
\vspace*{-8cm}
\caption{Geometry of the quantum well}
\label{geom}
\end{figure}

To begin with, we fix the geometry of the quantum well as shown in
Fig.~\ref{geom}. The figure shows a cut perpendicular to the $y$-axis,
the well extending from $z=-L/2$ to $z=L/2$. The remaining space
$|z|>L/2$ is occupied by the barrier material. The material parameters
of well and barrier are denoted by unprimed and primed characters,
respectively. To achieve a compact nomenclature, we define
$z$-dependent expressions, e.g.  \begin{equation} m(z) :=
m{\theta}(L/2-|z|) + m'{\theta}(|z|-L/2), \end{equation} for the mass
of a particle; $\theta(z)$ is the familiar step function.  As indicated
in the figure, we assume translational invariance within the $x$-$y$
plane; consequently, the corresponding components of the total momentum
are conserved.

To set up a model Hamiltonian, we proceed as described before.
Introducing center-of-mass and relative coordinates for the motion in
the $x$-$y$ plane, the corresponding center of mass part can be
eliminated by projection on the subspace of total momentum zero. We are
left with the Hamiltonian

\begin{eqnarray} \label{model1}
H &:=& {(2{\mu}_{\perp}(z_e,z_h))}^{-1}\vec{p}_{\perp}^{\,2} +
    p_{z,e}{(2{m_e}(z_e))}^{-1}p_{z,e} + p_{z,h}{(2{m_h}(z_h))}^{-1}p_{z,h}
    \nonumber \\
    && + V_e(z_e) + V_h(z_h) + V_{eh}(\vec{r}_{\perp}, z_e, z_h).
\end{eqnarray}
Here, the indices $e,h$ characterize  electron and hole, respectively;
$z_e$, $z_h$, $p_{z,e}$, and $p_{z,h}$ are the $z$-components of position
and momentum, ${m_e}(z)$ and ${m_h}(z)$ the corresponding masses. For the
relative coordinate in the $x$-$y$ plane we use the notation
$\vec r_{\perp} := \vec r_{\perp ,e} -\vec r_{\perp, h}$,
$\vec{p}_{\perp}$ denotes the corresponding momentum, and
\begin{equation}
({\mu}_{\perp}(z_e,z_h))^{-1} := (m_{\perp,e}(z_e))^{-1} +
(m_{\perp,h}(z_h))^{-1}
\end{equation}
is a generalized reduced mass. The confining potentials for electron and
hole are $V_e(z_e)$ and $V_h(z_h)$. These potentials are assumed to be
finite rectangular wells of the width $L$:
\begin{eqnarray}
\displaystyle V_i(z_i) := \left\{
\begin{array}{rcc}
 0, & \mbox{if} & |z_i| \leq L/2 \ , \\
 V_i, & \mbox{if} & |z_i| \geq L/2 \ ,
\end{array} \right.
\end{eqnarray}
where $i=e,h$. Finally, we have to specify $V_{eh}(\vec{r}_{\perp},
z_e,z_h)$, that is the potential energy of electron and hole. To do so,
we apply Poisson's equation to find the electrostatic potential of one
of the point charges (e.g. that of the hole) under the geometrical
conditions of Fig.~\ref{geom}. At the interfaces, the familiar
continuity conditions of electrodynamics have to be fulfilled. If we
assume this potential to be known, its product with the electron charge
yields the potential energy of interest. We note that the required
solution of Poisson's equation can conveniently be set up as follows.
At first, one performs a Fourier transform of the potential with
respect to the coordinates $x$ and $y$. One will realize that the
remaining differential equation with respect to $z$ can be treated
directly (the solutions are exponentials). In fact, this is the method,
which was used by Fomin and Pokatilov \cite{Fomin} under more general
circumstances. For the present geometry, their formulae can be somewhat
simplified, leading to an explicit result for the potential energy
which can be found in a paper of Kumagai and Tagakahara
\cite{Tagakahara}. These authors use the method of image-charges and
arrive at a power-series expansion of $V_{eh}$, the expansion parameter
being

\begin{equation} \label{q}
q:= {\varepsilon - \varepsilon' \over \varepsilon + \varepsilon'}.
\end{equation}
The zero'th order term $V_{eh}^{(0)}$ of the power series is the
familiar Coulomb expression

\begin{equation} \label{Coulomb1}
V_{eh}^{(0)}(\vec{r}_{\perp}, z_e, z_h) :=
- {\bar{e}^2\over{\varepsilon{\,r}}},
\end{equation}
the dielectric constant being that of the well material and
$\bar{e}$ denotes
the electron charge, divided by $\sqrt{4\pi{\varepsilon}_0}$ (SI units). The
higher-order terms constitute the so-called image potential. For
further use, we note the $q$-linear contribution explicitly:

\begin{equation} \label{image1}
V_{eh}^{(1)}(\vec{r}_{\perp}, z_e, z_h) :=
- {q \bar{e}^2\over\varepsilon}
\left(\frac{{\Theta}_{\gg}+{\Theta}_{\ll}}{r} +
 \frac{1-{\Theta}_{\gg}}{r_{-}} + \frac{1-{\Theta}_{\ll}}{r_{+}}\right).
\end{equation}
In this equation, we defined the distances $r$, $r_{\pm}$ and the
${\Theta}$-factors as follows:

\begin{eqnarray} \label{distance}
r &:=& \sqrt{r_{\perp}^{\,2}+(z_e - z_h)^2}, \nonumber \\
r_{\pm} &:=& \sqrt{r_{\perp}^{\,2}+(z_e + z_h \pm L)^2},
\end{eqnarray}
and
\begin{eqnarray}  \label{theta}
{\Theta}_{\gg} &:=& {\theta}(z_{e} -L/2) + {\theta}(z_{h} -L/2), \nonumber\\
{\Theta}_{\ll} &:=& {\theta}(-z_{e} -L/2) + {\theta}(-z_{h} -L/2).
\end{eqnarray}
We add three remarks concerning these formulae.

The reader will notice that all terms of $V_{eh}^{(1)}$ are of Coulomb
type, the corresponding denominators being partially displaced.  It is
important to realize that this property remains valid for all
higher-order contributions (again, see Ref. \onlinecite{Tagakahara}).
Consequently, we induce an error of order $q^2$, if we replace the full
expression for $V_{eh}$ by $V_{eh}^{(0)} + V_{eh}^{(1)}$. As $q$ is
typically of the order 0.1, this simplification is reasonable and will
be used in the remainder of this paper.

The second remark is concerned with an ambiguity in the power-series
expansion of $V_{eh}$. Because of Eq.~(\ref{q}), we may replace
$\varepsilon$ by ${\varepsilon'}(1+q)/(1-q)$. Inserting this equality
into Eqs. (\ref{Coulomb1}) and (\ref{image1}), we find terms of zero'th
and first order as follows:

\begin{equation} \label{Coulomb2}
\tilde{V}_{eh}^{(0)}(\vec{r}_{\perp}, z_e, z_h) :=
- {\bar{e}^2\over{\varepsilon' r}},
\end{equation}

\begin{equation} \label{image2}
\tilde{V}_{eh}^{(1)}(\vec{r}_{\perp}, z_e, z_h) :=
- {q \bar{e}^2\over\varepsilon'}
\left(\frac{{\Theta}_{\gg}+{\Theta}_{\ll}-2}{r} +
 \frac{1-{\Theta}_{\gg}}{r_{-}} + \frac{1-{\Theta}_{\ll}}{r_{+}}\right).
\end{equation}
These expansions are clearly equivalent, and will be used to our
advantage in the next section.

Finally, we mention that some authors (see, e.g.,
Refs.~\onlinecite{Andreani,Tagakahara,Zimmermann}) include self-energy
terms in $H$, which are also due to a dielectric mismatch.  Formally
these appear if one considers an electron- and a hole-inhomogeneity in
Poisson's equation and inserts the total electrostatic energy into $H$.
The interaction part of this energy is the one we used above (i.e.
$V_{eh}(\vec{r}_{\perp},z_e,z_h)$), the diagonal parts produce the
self-energy terms $\Sigma (z_e)$ for the electron (i.e.
$-V_{eh}(0,z_e,z_e)/2$, evaluated for $\vec{r}_{\perp}=0$ and
$z_h=z_e$, the bulk Coulomb singularity being subtracted) and
$\Sigma(z_h)$ for the hole. Inspection of Eq.~(\ref{image1}) shows that
$\Sigma (z)$ consists of one-dimensional Coulomb potentials, which
exhibit a singularity on the interfaces $z=+L/2$ and $z=-L/2$ and
vanish if $q$ vanishes. Summarizing, the confinement potentials are
claimed to be changed. We doubt that modifications of the confinement
potentials can be motivated this way and refer to the early
quantum-mechanical debate on the hydrogen problem with (or without)
self-energy corrections (e.g. Tomonaga \cite{Tomonaga}). Consequently,
we did not include such corrections here. We agree, however, with the
statement made in Ref.~\onlinecite{Zimmermann} that a correct
microscopic description of an exciton in a quantum well should contain
(well behaved) polarization-induced modifications of the rectangular
confinement potentials.  It should be mentioned that if we modify the
confinement potentials in the way indicated above, then the corrections
to the final binding energies presented in Sec.~IV only appear in
the vicinity of the peak being not larger than 0.5 meV.

It will prove useful to introduce dimensionless variables. For the
excitonic system under consideration, appropriate units of length and
energy are Bohr radius and Rydberg energy. Considering the well
material, these are

\begin{equation}  \label{units}
a_B := {\hbar^2\varepsilon\over \mu_{\perp}\bar{e}^2}, \quad
Ry :={\mu_{\perp}\bar{e}^4\over 2\varepsilon^2\hbar^2}.
\end{equation}
Returning to Eqs. (\ref{model1}), (\ref{Coulomb1}), (\ref{image1}), we
replace $\vec{r}$ by $a_B \vec{r}$, introduce the dimensionless Hamiltonian

\begin{equation}
h:=H/Ry
\end{equation}
and find the expression

\begin{eqnarray} \label{model2}
h &=& -{\mu_{\perp} \over  \mu_{\perp}(z_e,z_h)}{\vec\nabla}_{\perp}^2 -
{\partial \over \partial z_e}{\mu_{\perp} \over {m_e}(z_e)}{\partial \over
\partial z_e} -
{\partial \over \partial z_h}{\mu_{\perp} \over {m_h}(z_h)}{\partial \over
\partial z_h}  \nonumber\\[2mm]
&& - {2\over r}
- 2q \left(\frac{{\Theta}_{\gg}+{\Theta}_{\ll}}{r} +
 \frac{1-{\Theta}_{\gg}}{r_{-}} + \frac{1-{\Theta}_{\ll}}{r_{+}}\right)
\nonumber \\[2mm]
&& + U_e(z_e) + U_h(z_h).
\end{eqnarray}
All variables are now dimensionless. The confinement potentials read as
follows;

\begin{eqnarray} \label{confine1}
\displaystyle U_i(z_i) := \left\{
\begin{array}{rcc}
 0, & \mbox{if} & |z_i| \leq l/2, \\
 U_i, & \mbox{if} & |z_i| \geq l/2,
\end{array} \right.
\end{eqnarray}
where we introduced

\begin{equation}  \label{confine2}
U_i:= {V_i \over Ry}, \quad l:={L\over a_B}.
\end{equation}
The theta factors are defined as in Eq.~(\ref{theta}) with the
exception that $L$ is replaced by $l$.

We shall need a second dimensionless version of Hamiltonian
(\ref{model1}), which is based on the Bohr radius $a_B'$ and the
Rydberg energy $Ry'$ of the barrier material as well as the formulae
(\ref{Coulomb2}), (\ref{image2}) for the electron-hole potential
energy.  Introducing

\begin{equation}
h':=H/Ry' \ \,
\end{equation}
one arrives at

\begin{eqnarray}  \label{model3}
h' &=& - {\mu_{\perp}'\over \mu_{\perp}(z_e,z_h)} {\vec \nabla}_{\perp}^2 -
{\partial \over \partial z_e}{\mu_{\perp}' \over {m_e}(z_e)}{\partial \over
\partial z_e} -
{\partial \over \partial z_h}{\mu_{\perp}' \over {m_h}(z_h)}{\partial \over
\partial z_h}  \nonumber\\[2mm]
&& - {2\over r}
- 2q \left(\frac{{\Theta}_{\gg}'+{\Theta}_{\ll}'-2}{r} +
 \frac{1-{\Theta}_{\gg}'}{r_{-}} + \frac{1-{\Theta}_{\ll}'}{r_{+}}\right)
\nonumber \\[2mm]
&&+ U_e'(z_e) + U_h'(z_h).
\end{eqnarray}

The theta factors are defined as in Eq.~(\ref{theta}) with the
exception that $L$ has to be replaced by $l':=L/a_b'$. In analogy, the
confinement potentials $U'$ can be taken from Eqs. (\ref{confine1}) and
(\ref{confine2}), if one replaces the unprimed material parameters by
primed ones.

\section{Upper and lower bounds for the exciton binding energy}

Because of the absence of translation symmetry, the Hamiltonian $H$ (or
$h$, $h'$, respectively) cannot be treated analytically. Nevertheless,
we can fix quantitative properties of the binding energy $E_b$ by means
of rigorous bounds, which will be derived in the following sections.
Our strategy will be to discuss firstly the ground-state energy $E_0$.
The reason is that we can directly prove bounds for $E_0$. The binding
energy $E_b$, in turn, is related to $E_0$ by the equation

\begin{equation} \label{binding}
E_b=E_{cont}-E_0,
\end{equation}
where $E_{cont}$ denotes the energy of the continuum edge. In our case,
$E_{cont}$ is the sum of the lowest well energies of electron and hole,
any correlation being neglected. For further use we note the explicit
formulae

\begin{equation}
E_{cont}/Ry = e_{0,e} + e_{0,h} ,
\end{equation}
where $e_{0,i}\ (i=e,h)$ is the ground-state eigenvalue of the
one-dimensional well Hamiltonian

\begin{equation}  \label{Hwell}
h_i:= - {\partial \over \partial z}{\mu_{\perp} \over  {m_i}(z)}{\partial \over
\partial z} + U_i(z),
\end{equation}
part of the total Hamiltonian (\ref{model2}). The eigenvalue $e_{0,i}$
is implicitly defined by the equation

\begin{equation}  \label{Ewell}
e_{0,i}={4\mu_{\perp}\over m_i\,l^2 }
\arcsin^2{\sqrt{1-e_{0,i}/U_i}\over \sqrt{1+({m'_i}/{m_i}-1)e_{0,i}/U_i}}.
\end{equation}
Due to Eq.~(\ref{binding}), an upper (lower) bound for $E_0$ yields a
lower (upper) bound for $E_b$.

\subsection{Lower bounds for the binding energy}

We use a variational approach, to derive an upper bound on $E_0$.  The
trial wave function is

\begin{equation} \label{trial}
\Psi_{\alpha,\lambda}(\vec{r}_{\perp},z_e,z_h):=\Phi_e(z_e) \Phi_h(z_h)
e^{-\alpha\sqrt{r_{\perp}^2+\lambda z^2}} \hfill ,
\end{equation}
where $z:=z_e-z_h$ and $\Phi_i\ (i=e,h)$ is the ground-state eigenfunction
of $h_i$ (see Eq.~(\ref{Hwell})). Clearly, the factors $\Phi_e$ and $\Phi_h$
serve to incorporate the single-particle well behavior.


Choosing an exponential envelope in Eq.~(\ref{trial}), we account for
two important aspects. Firstly, the con\-fine\-ment-induced change of
the effective Bohr radius (parameter $\alpha$), and, secondly, the
quenching of the wave function in $z$-direction (parameter $\lambda$).
Both aspects are particularly important if one approaches the quasi
two-dimensional case, where the width of the layer becomes considerably
smaller than the Bohr radius of the exciton. We remark that the
envelope can correctly reproduce the ultimate limits of a free exciton
ground-state in two or three dimensions. Subsequently, the trial
function is quite flexible and we expect the variational inequality

\begin{eqnarray}
{E_0\over Ry} \leq \min_{\lambda,\alpha}
{<\Psi_{\alpha,\lambda}|h|\Psi_{\alpha,\lambda}> \over
<\Psi_{\alpha,\lambda}|\Psi_{\alpha,\lambda}>}
\end{eqnarray}
to be effective. For the binding energy $E_b$ we derive

\begin{eqnarray} \label{bound1}
{E_b\over Ry} \geq e_{0,e}+e_{0,h} - \min_{\lambda,\alpha}
{<\Psi_{\alpha,\lambda}|h|\Psi_{\alpha,\lambda}> \over
<\Psi_{\alpha,\lambda}|\Psi_{\alpha,\lambda}>} .
\end{eqnarray}

Note that the minimum of the right-hand side can be calculated for
$q=0$, as we have linearized $h$ with respect to $q$.

\subsection{Upper bounds for the binding energy}

In this subsection, we provide a class of lower bounds for $E_0$.  To
do so, we assume that the barrier masses $m_e', m_h'$ are not smaller
than the corresponding well masses $m_e, m_h$. This is the case for
$Ga_{1-x}Al_{x}As/GaAs/Ga_{1-x}Al_{x}As$, which will be used as an
example. Recalling formula (\ref{model3}) for the Hamiltonian under
consideration, one verifies by direct inspection that the following
inequality is true in the sense of operators:

\begin{equation}
h'\geq \bar{h'},
\end{equation}
where

\begin{eqnarray} \label{h2}
\bar{h'} &=&:
- {\vec \nabla}_{\perp}^2
- {\mu_{\perp}' \over m_e'}{\partial^2 \over \partial z_e^2}
- {\mu_{\perp}' \over m_h'}{\partial^2 \over \partial z_h^2}  \nonumber \\
&& - {2\over r} - 2q \left(\frac{1}{r_{-}} + \frac{1}{r_{+}} \right)+
U_e'(z_e) + U_h'(z_h).
\end{eqnarray}
We stress that the above assumption $m_i' \geq m_i$ is not crucial. If the
contrary was true, we would find an analogous inequality with respect to
Hamiltonian (\ref{model2}).

To provide a lower bound to the spectrum of $\bar{h'}$, we split the
Hamiltonian (\ref{h2}) into four tractable parts. Assuming the
corresponding ground-state energies to be known, their sum will be a
lower bound to $E_0/Ry'$. Let us consider the following decomposition:

\begin{eqnarray} \label{h2split}
\bar{h'}&=& h'_e + h'_h +h'_c + h'_{im}, \nonumber \\[2mm]
h'_i &=&: -(1-x_i){\mu_{\perp}'\over m_i'}{\partial^2\over \partial z_i^2} +
U'_i(z_i), \qquad i=e,h, \nonumber \\[2mm]
h'_c &=&: - y {\vec \nabla}_{\perp}^{\,2} -
x_e y_e {\mu_{\perp}'\over m_e'}{\partial^2\over \partial z_e^2} -
x_h y_h {\mu_{\perp}'\over m_h'}{\partial^2\over \partial z_h^2} -
{2\over r}, \nonumber \\[2mm]
h'_{im} &=&: - (1-y) {\vec \nabla}_{\perp}^{\,2} -
x_e (1-y_e) {\mu_{\perp}'\over m_e'}{\partial^2\over \partial z_e^2}
\nonumber \\
&& - x_h (1-y_h) {\mu_{\perp}'\over m_h'}{\partial^2\over \partial z_h^2}
- 2q \left(\frac{1}{r_{-}} + \frac{1}{r_{+}} \right).
\end{eqnarray}
Here, $x_i, y_i, y$ are parameters with values in the interval $[0,1]$
but otherwise arbitrary, which can then be chosen to lift the lower
bound as much as possible. The exciton ground-state energy $E_0$
fulfills the inequality

\begin{equation}  \label{lbound}
E_0/Ry' \geq e'_e + e'_h + e'_c + e'_{im},
\end{equation}
where the four terms on the right-hand side are the ground-state
eigenvalues of the four parts of $\bar{h'}$, defined in
Eq.~(\ref{h2split}).  The eigenvalues $e'_i$ can be derived from
Eq.~(\ref{Ewell}); removing the mass mismatch and rescaling the
particle mass appropriately, we find

\begin{equation}   \label{well1}
e'_i = 4(1-x_i){\mu_{\perp}'\over m_i'l'^2}
\arcsin^2 \sqrt{1-e'_i/U'_i}.
\end{equation}
Concerning $e'_c$, we transform the corresponding Hamiltonian $h'_c$;
introducing center-of-mass and relative coordinates instead of $z_e$
and $z_h$, we can separate the center-of-mass part. Performing an
appropriate scaling transformation with respect to ${\vec r}_{\perp}$
and $z=z_e -z_h$, we arrive at the following equations;

\begin{equation} \label{lbound2}
e'_c = {1\over y} \mbox{\rm inf spec} \left\{-\vec \nabla^{\,2} -
{2\over \sqrt{r_{\perp}^2 + Az^2}}\right\} =:{1\over y}e(A) ,
\end{equation}
where

\begin{equation} \label{anisotropy1}
A = {\mu'_{\perp}\over \mu'_{\parallel}}\,{x_e y_e m'_h +
x_h y_h m'_e\over y(m'_e+m'_h)}, \quad
\mu'_{\parallel} = {m'_e m'_h \over m'_e+m'_h}.
\end{equation}
Clearly, $e(A)$ is (in units of $Ry'$) the ground-state energy of an
anisotropic Coulomb system. Before we comment on this, we turn to the
last term in inequality (\ref{lbound}), namely $e'_{im}$. We transform
the corresponding operator $h'_{im}$ in two steps. Firstly, we replace
$z_h$ by $-z_h$, secondly, we introduce center-of-mass and relative
coordinates instead of $z_e$ and $z_h$ as was done above. We find

\begin{equation}    \label{lbound3}
e'_{im} = \mbox{\rm inf spec}\left\{- (1-y) {\vec \nabla}_{\perp}^{\,2}
- \mu'_{\perp} \left[{x_e (1-y_e) \over m'_e} + {x_h(1-y_h)\over m'_h}\right]
{\partial^2\over \partial z^2}
- 2q \left[\frac{1}{R_{-}} + \frac{1}{R_{+}} \right]\right\},
\end{equation}
where $R_{+}$ and $R_{-}$ are derived from $r_{+}$ and $r_{-}$ by
replacing $z_e + z_h$ by the relative coordinate $z$. Splitting the
right-hand side of expression (\ref{lbound3}) again into two parts,
containing the terms $1/R_{-}$ and $1/R_{+}$ separately, we may shift
the variable $z$ by $+l'$ and by $-l'$, respectively.  We obtain pure
Coulomb potentials in both cases. Assuming their solutions to be known,
we are finally lead to the inequality

\begin{equation}    \label{lbound4}
e'_{im} \geq \mbox{\rm inf spec} \left\{ - (1-y) {\vec \nabla}_{\perp}^{\,2}
- \mu'_{\perp} \left[{x_e (1-y_e) \over m'_e} + {x_h(1-y_h)\over m'_h}\right]
{\partial^2\over \partial z^2}
- \frac {4q}{r} \right\}.
\end{equation}

Performing an appropriate scaling of variables, we can reduce the
right-hand side once more to the anisotropic Coulomb problem. In
comparison with Eq.~(\ref{lbound2}), the parameters are changed:

\begin{equation}  \label{lbound5}
e'_{im} \geq {4q^2 \over 1-y} \mbox{\rm inf spec} \left\{-\vec \nabla^{\,2} -
{2\over \sqrt{r_{\perp}^2 + A'z^2}}\right\},
\end{equation}
the anisotropy parameter now being

\begin{equation} \label{anisotropy2}
A' = {\mu'_{\perp}\over \mu'_{\parallel}}\,{x_e (1-y_e) m_h +
x_h (1-y_h) m_e\over (1-y)(m_e+m_h)}.
\end{equation}

We summarize our results for $e'_c$, $e'_{im}$, and $E_0/Ry'$. To
evaluate the relations (\ref{lbound}), (\ref{lbound2}) and
(\ref{lbound5}), we need an exact expression for the ground-state
eigenvalue $e(A)$ of the anisotropic Coulomb problem or, at least,
a corresponding lower bound. Assuming $e(A)$ to be known, the above
considerations prove the following inequality:

\begin{equation} \label{lbound6}
{E_0\over Ry'} \geq \max_{x_e,x_h,y_e,y_h,y}
\left[e'_e + e'_h + {1\over y} e(A) + {4q^2\over (1-y)} e(A')\right].
\end{equation}
Finally, we arrive at a result for the binding energy, which complements
the relation (\ref{bound1});

\begin{equation}  \label{bound2}
{E_b \over Ry'} \leq -\max_{x_e,x_h,y_e,y_h,y}
\left[(e'_e - e_{0,e}) + (e'_h - e_{0,h}) + {1\over y} e(A) +
{4q^2\over (1-y)} e(A')\right].
\end{equation}

\noindent
To the best of our knowledge, an exact analytical equation for $e(A)$
is not available up to now. However, involved variational calculations
were performed by several authors; we refer to the paper by Gerlach and
Pollmann \cite{Gerpol}, the references therein and to our brief summary
in the Appendix.  The optimal upper bound on $e(A)$, which is presented
in Ref. \onlinecite{Gerpol}, will be denoted as $e_{GP}(A)$. Numerical
studies indicate that $e_{GP}(A)$ deviates from $e(A)$ only on a
one-percent scale, although a rigorous estimate is admittedly missing.
We mention that the quoted paper also contains additional lower bounds
for $e(A)$ which might be used to evaluate the preceeding inequality.
In this paper we shall add another lower bound (again, see the
Appendix). Unfortunately, the overall quality of all these lower bounds
is not sufficient. Therefore, we will usually insert the numerical
approximation $e_{GP}(A)$ instead of $e(A)$.

\section{Results and discussion}

\subsection{Absence of mass- and dielectric mismatch}

We used this simplified model to test the efficiency of the above
bounds and to illustrate the dimensional aspect. The system
is interesting on it's own and was already discussed in the literature
(see Refs. \onlinecite{Andrea,Bajaj}). Here, it will prove useful to
understand the well-width dependence of the binding energy on a
quantitative basis.

One can easily specify the general results from above for the
discussion of the present case. Equalizing all primed and unprimed
material parameters, we have $q=0$ and $h=h'$. We simplify even further
by assuming $m_e=m_h$, $\mu_{\perp} = \mu_{\parallel} = {m_e}/2$, and
$U_e'=U_h'=U$. Turning to inequality (\ref{bound2}) and recalling its
derivation, we have to choose $y_e = y_h = y = 1$ (no image potential).
Consequently, the anisotropy parameter (see Eq.~(\ref{anisotropy1})) is
fixed as $A=(x_e+x_h)/2$. Summarizing at this stage, we derive from
relation (\ref{bound2}):

\begin{equation}  \label{bound3}
{E_b \over Ry} \leq -\max_{x_e,x_h}
\left[(e_e - e_{0,e}) + (e_h - e_{0,h}) + e(A)\right],
\end{equation}
where $e_i$ (see Eq.~(\ref{well1}) is now the solution of the equation

\begin{equation}   \label{well2}
e_i = {2(1-x_i) \over l^2}\arcsin^2 \sqrt{1-e_i/U},
\end{equation}
and $e_{i,0}$ can be derived from $e_i$, if we let $x_i=0$. The
eigenvalue $e(A)$ was defined in Eq.~(\ref{lbound2}). We shall now
treat the cases of an infinite and a finite well. They will be shown to
have a qualitatively different behavior in the narrow-well regime, as
indicated in the Introduction.

For the limiting case of an infinite well we can simplify the inequality
(\ref{bound3}) as follows:

\begin{equation}  \label{bound4}
{E_b \over Ry} \leq -\max_{A} \left[-A{\pi^2 \over l^2} + e(A)\right].
\end{equation}
To proceed, we employ another inequality, namely  $e(A) \geq -4/(1+3A)$,
which is derived in the Appendix. In this case, we are lead to an
analytical result for $E_b$:

\begin{eqnarray} \label{bound5}
\displaystyle {E_b \over Ry} \leq \left\{
\begin{array}{rcl}     \displaystyle
 4, & \mbox{if} &
\displaystyle l \leq {\pi \sqrt{3} \over 6} \approx 0.91, \\[4mm]
\displaystyle  {4\pi\sqrt{3}\over 3l} - {\pi^2\over 3l^2}, & \mbox{if} &
0.91 \leq l \leq 3.63, \\[4mm]
\displaystyle 1+{\pi^2\over l^2}, & \mbox{if} &
\displaystyle l \geq {2\pi\sqrt{3}\over 3} \approx 3.63.
\end{array} \right.
\end{eqnarray}

\begin{figure}
\epsfxsize=14cm \epsfbox{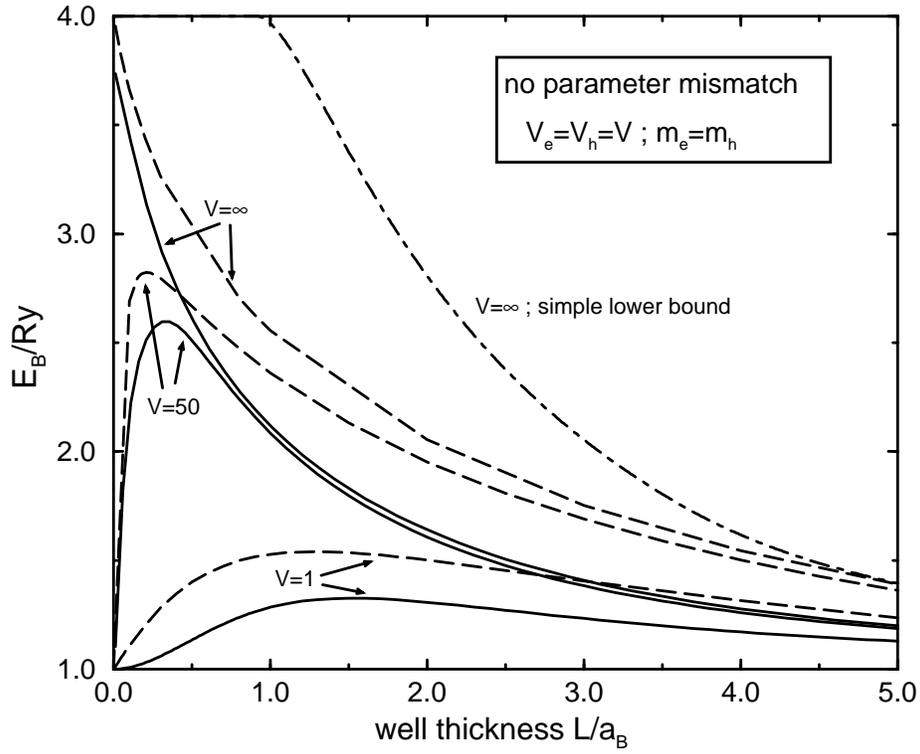}
\caption{Binding energy as function of the well thickness for different
potential heights, showing comparison of variational calculation (solid
lines) and lower bound method (dashed lines). The dashed-dotted line
shows the result for the simple analytical lower bound from the
Appendix.}
\label{exc1}
\end{figure}

The right-hand side of Eq.~(\ref{bound5}) is shown as dashed-dotted
curve in Fig.~\ref{exc1}.  One realizes that the required two- and
three-dimensional limits 4\,$Ry$ and 1\,$Ry$ are reproduced, if $l$
tends to $0$ and $\infty$, respectively.  Unfortunately, the overall
quality of this bound is rather poor. Inserting the result $e_{GP}(A)$
for $e(A)$ as indicated in the preceding section, the upper bound on
$E_b$ is drastically lowered as shown in the figure (dashed line for
$V=\infty$). In addition, Fig.~\ref{exc1} contains a lower bound for
the binding energy (solid curve for $V=\infty$), which is based on
inequality (\ref{bound1}).

Before we discuss the results in greater detail, we consider the
finite-well case. Then, the inequality (\ref{bound3}) has to be
evaluated without further simplifications. Apart from the limiting
cases $l\to 0$ and $l\to \infty$, an analytical discussion is not
possible. The small-width case, however, is interesting on its own. For
$l \to 0$, the solution for $e_i$ is

\begin{equation}
e_i = U[1-Ul^2/(2-2x_i)+O(l^3)].
\end{equation}

\noindent
Inserting this expression in inequality (\ref{bound3}) and utilizing
$e(A) \geq -4/(1+3A)$ again, one finds the result

\begin{equation} \label{bound6}
1 \leq {E_b \over Ry} \leq 1 + \sqrt{3}Ul + O(l^2),
\quad \mbox{\rm if} \quad l \to 0,
\end{equation}

\noindent
which was mentioned in the abstract as well as the Introduction.  For
finite values of $l$, we proceed as in the numerical treatment of the
infinite well. Replacing $e(A)$ by $e_{GP}(A)$, we find the upper
bounds, which are shown in Fig.~\ref{exc1} (dashed curves).  Again, the
figure  includes the corresponding lower bounds (solid curves).

Clearly, the most significant attribute of these curves is the
appearance of a maximum of $E_b$ as function of the well width $L$, if
the height $V$ of the well is finite. There exists a unique relation
between $V$ and the position $L_{max}$ of the maximum; for larger
values of $V$ $L_{max}$ becomes smaller.  All curves exhibit the
expected asymptotic behavior. For finite (infinite) $V$, the large-$L$
limit of the binding energy is $1\,Ry$, the small-$L$ limit $1\,Ry$
($4\,Ry$). Last but not least, the shape of the upper and lower bound
is very similar, the relative deviation not exceeding 0.2. We are sure
that the deviation as such is mostly due to the inaccuracy of the upper
bound on the binding energy. We recall that the lower bound for $E_b$
is based on a variational upper bound on the ground-state energy,
whereas the upper bound needs an accurate lower bound for $E_0$ as
input; as usual, this part of the task is the more difficult one.

\subsection{Finite rectangular quantum well in a heterostructure}

In this part we apply our theory to a single-well structure, which
exhibits a mismatch of both masses and dielectric parameters. As far as
specific material data are concerned, we use those of
$Ga_{1-x}Al_{x}As/GaAs/Ga_{1-x}Al_{x}As$. We stress, however, that a
direct comparison of our results with experimental data is limited by
the fact that the present theory is clearly incomplete. An obvious
shortcoming is that, for example, the effects of valence-band
degeneracy, spin-orbit coupling, and exchange interaction are not
included. Our intention was to analyze the implications of the well
structure, in particular the inability to separate the center-of-mass
and relative coordinates, as accurate as possible in order to have a
well-defined basis for further improvements.

For the material parameters, we refer to the tables of
Landolt-B\"ornstein \cite{LB} and the work of Winkler \cite{Winkler}
(see also references therein). For $GaAs$, we used $\varepsilon=12.53,\
m_e=0.067m_0$, furthermore $m_h=0.090m_0,\ \mu_{\perp}=0.051m_0,\
Ry=4.418\ \mbox{\rm meV},\ a_B=130.21\ \AA$ for the light hole, and
$m_h=0.377m_0,\ \mu_{\perp}=0.042m_0,\ Ry=3.638\ \mbox{\rm meV},\
a_B=158.12\ \AA$ for the heavy hole. For $AlAs$, the corresponding
parameters are $\varepsilon=10.06\, ,\,\ m_e=0.150m_0\, ,\,
m_h=0.208\,(0.478)m_0 \, ,\, \mu_{\perp}=0.106\,(0.093)m_0\, ,\,
Ry=14.2\,(12.5)\ \mbox{\rm meV}\, ,\, a_B=50.2\,(57.2)\ \AA$ for the
light (heavy) hole. The material parameters for $Ga_{1-x}Al_{x}As$ are
normally described by linear interpolation formulae, for instance,
$\varepsilon' = 12.53 (1-x) + 10.06 x$, with one notable  exception;
the difference $\Delta E_g$ of the gap-energies for $x=0$ and $x>0$ is
fitted as\cite{Winkler}

\begin{equation}
\Delta E_g = (1.087x + 0.438x^2)\ {\mbox eV}.
\end{equation}

The precise partition of $\Delta E_g$ on the electronic part $V_e$ and
the hole part $V_h$ has been critically discussed; to the best of our
knowledge, no previous general solution exists. We used the empirical
relation $V_e = 0.65\Delta E_g$ and $V_h = 0.35\Delta E_g$ (again, see
Ref.~\onlinecite{Winkler}), but also indicate the consequences of
modification to this assumption.

\begin{figure}
\epsfxsize=14cm \epsfbox{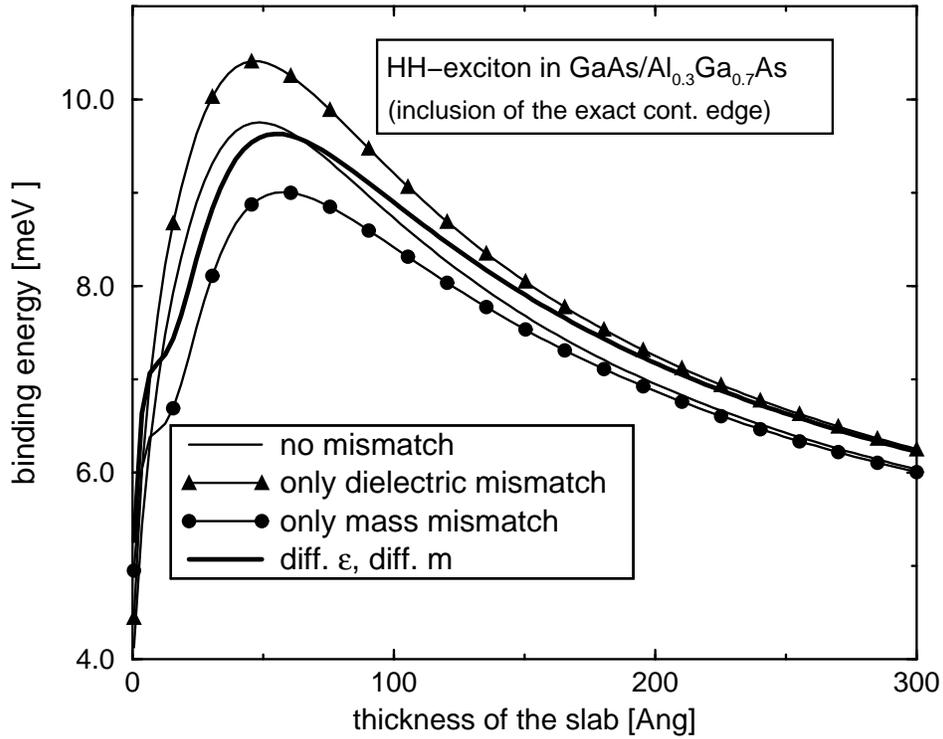}
\caption{Binding energy of a HH-exciton in
$Al_{0.3}Ga_{0.7}As/GaAs/Al_{0.3}Ga_{0.7}As$; the effect of dielectric
and mass mismatch is shown separately by the triangle and circle
symbols, respectively. The thick solid line illustrates the result of
both contributions.}
\label{exc2}
\end{figure}

We discuss now our results. To begin with, we present lower bounds for
the binding energy according to relation (\ref{bound1}). In
Fig.~\ref{exc2} we depict the influence of the parameter mismatches on
the heavy-hole (HH) exciton of $Al_xGa_{1-x}As/GaAs/Al_xGa_{1-x}As$ for
$x=0.3$. The thin solid curve describes the hypothetical case of equal
masses and dielectric constants in well and barrier; it may be viewed
as a reference line. Switching on the discontinuity of the band masses
only (circles), the bound for the binding energy is lowered. We are
confident that this is an artifact of our variational treatment. On the
one hand, the energy of the continuum edge is exact, the effects of the
mass mismatch being fully incorporated; on the other hand, the
variational bound on the ground-state energy is, of course, an
approximation. The difference between the two underestimates the
influence of the mass mismatch. Considering only a dielectric mismatch,
we find the curve depicted by triangles.  One realizes that the
presence of image charges shifts the binding energy to higher values,
and this shift remains present for a wider range of the well width. It
is only for $L \gg a_B$ that the energy shift disappears. Finally, the
thick solid line summarizes all effects. In comparison to the reference
curve, the peak height and peak position is only slightly changed; in
our case, we find $L_{max}\sim 60\, \AA$ and a peak height of $9.5$
meV.

\begin{figure}
\epsfxsize=14cm \epsfbox{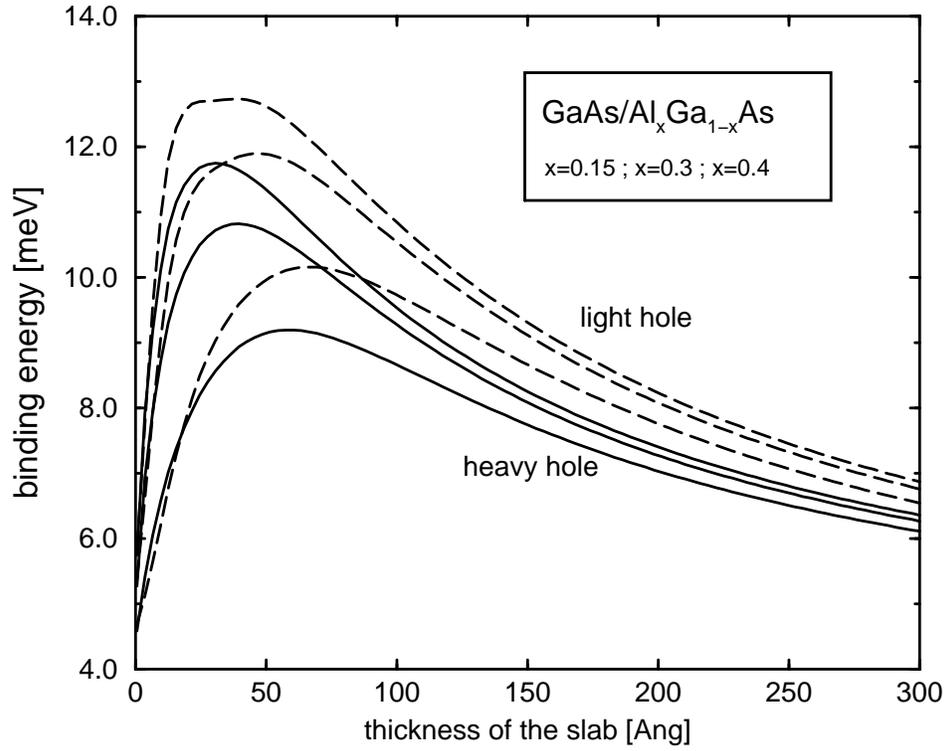}
\caption{Binding energies of the HH- and LH-exciton for different
$Al$-concentrations $x=0.15 , x=0.3 , x=0.4$. The variational approach
including parameter mismatch is applied. The lowest (highest) curves
belong to the smallest (largest) value of $x$.}
\label{exc3}
\end{figure}

Fig.~\ref{exc3} presents lower bounds for the binding energies of
heavy- and light-hole excitons in $Al_xGa_{1-x}As/GaAs/Al_xGa_{1-x}As$,
assuming different $Al$-concentrations $x=0.15,\ x=0.3,\ x=0.4$. The
dashed (solid) curves correspond to the light-hole (heavy-hole)
exciton. The lowest (highest) curve of every set belongs to the
smallest (largest) value of $x$. The observed sequence can easily be
explained. A higher $Al$-concentration causes a higher band-gap
difference and therefore higher potential wells. In addition, the peak
positions are found to decrease with increasing $Al$-concentration.
Again, the reason is clear; the higher the value of $x$, the smaller
the influence of the barrier. In fact, we expect to recover the trend
illustrated in Fig.~\ref{exc1}, which is indeed the case.  Above all,
we find a peak structure for $x>0.25$ and $L>30\,\AA$, in contrast to
the assertion of Andreani and Pasquarello \cite{Andreani}.

\begin{figure}
\epsfxsize=14cm \epsfbox{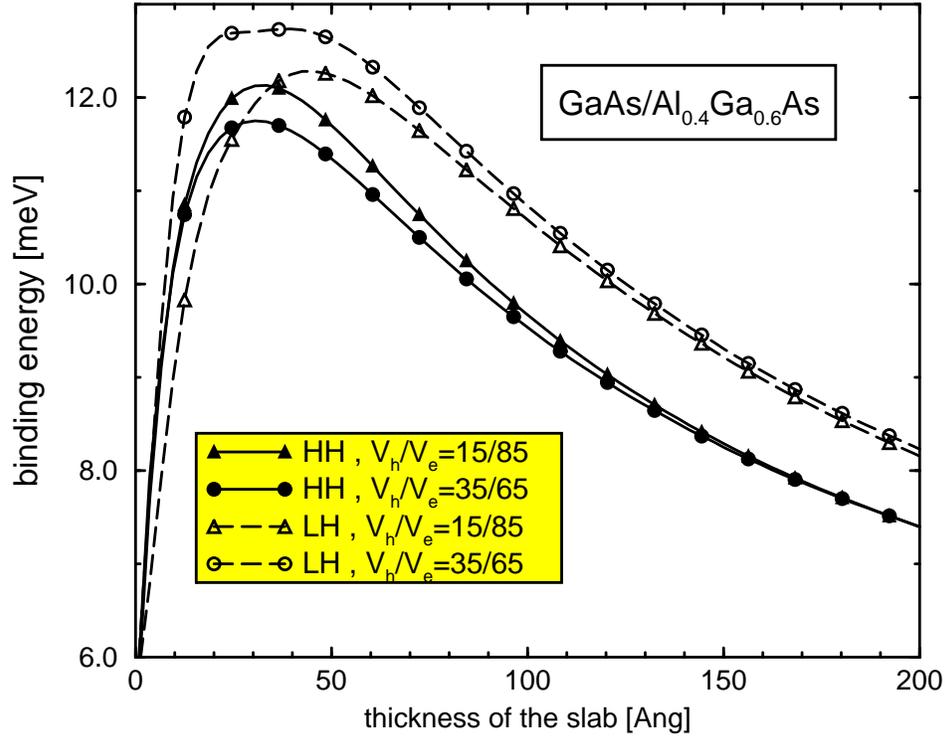}
\caption{Binding energies of the HH- and LH-exciton for different
valence band offsets in $Al_{0.4}Ga_{0.6}As/GaAs/Al_{0.4}Ga_{0.6}As$.
The valence band offsets are chosen either from Greene, Bajaj and
Phelps\protect\cite{Bajaj} (triangles) or from Andreani and
Pasquarello\protect\cite{Andreani} (circles).}
\label{exc4}
\end{figure}

We mentioned above that the actual values of $V_e$ and $V_h$ are
somewhat controversial. Therefore, we found it interesting to change
the fraction $V_h/V_e$. In Fig.~\ref{exc4} we compare lower bounds for
the binding energy, $V_h/V_e$ being chosen as $35/65$ (circles) and
$15/85$ (triangles), respectively; the latter value was used by Greene,
Bajaj and Phelps \cite {Bajaj}. Interestingly enough, the peak position
is not changed too much.

\begin{figure}
\epsfxsize=14cm \epsfbox{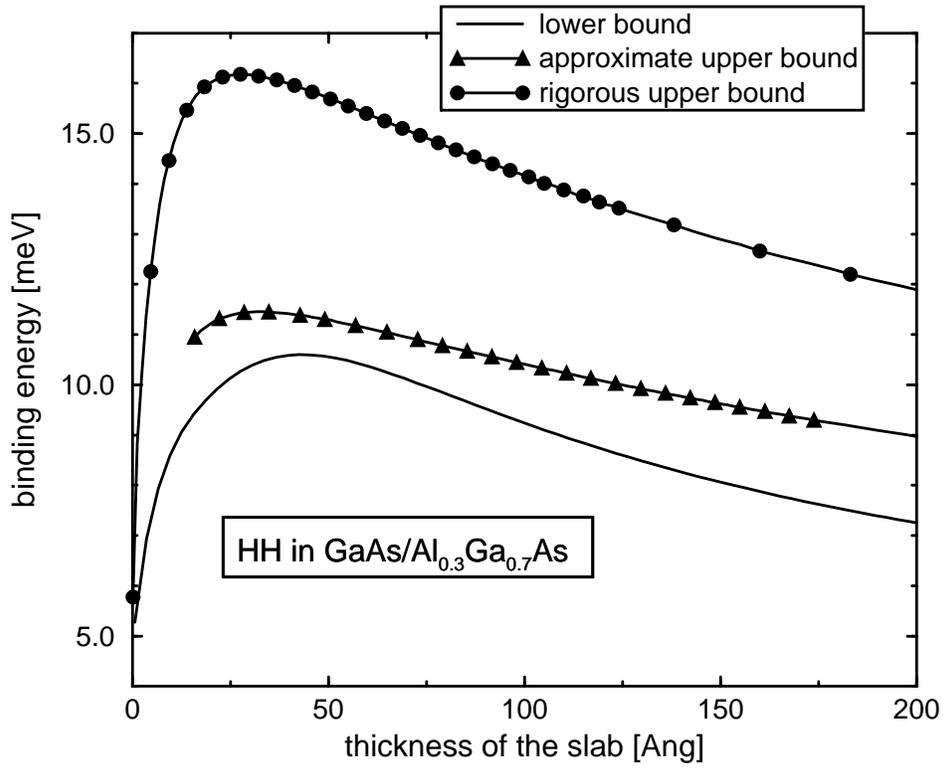}
\caption{Rigorous upper bounds (circles) and lower bounds (lines) in
comparison with approximate upper bounds for the exciton binding energy
in $Al_{0.3}Ga_{0.7}As/GaAs/Al_{0.3}Ga_{0.7}As$.}
\label{exc5}
\end{figure}

We shall now comment on the accuracy of our results. To do so, we have
chosen the HH-case of $Al_{0.3}Ga_{0.7}As/GaAs/Al_{0.3}Ga_{0.7}As$ as a
representative example. In Fig.~\ref{exc5} we contrast lower and upper
bounds for the binding energy according to
Eqs. (\ref{bound1}) and (\ref{bound2}). The shape of the curves and the
peak positions are nearly the same for both bounds, the peak heights,
however, differ disappointingly by a factor 1.5. Analyzing the
numerical data in detail, we realized two points: i) the adapted trial
function is well localized within the well for all values of $L \geq
L_{max}$ and even slightly below; ii) the upper bound for the binding
energy (or, equivalently, the lower bound for the ground-state energy)
grossly overestimates the mismatch. If we take  (i) for granted, also
for the exact wave function, we can simplify the peak problem from the
very beginning. If we calculate an expectation value of $h$ (see
Eq.~(\ref{model2})) for such a wave function, we can firstly omit all
$\Theta$-factors, and secondly replace all material parameters by well
parameters. Then, $h$ may be replaced by $\bar{h'}$ (see
Eq.~(\ref{h2})) with the important modification that all parameters in
$\bar{h'}$ have to be understood as unprimed ones, namely those of the
well. Calculating lower bounds in exactly the same manner as before, we
find the bounds which are denoted by approximate ones (triangles). Of
course, this curve has to be cut in the region of small well widths
when the requirement of prevailing localization inside the well is not
fulfilled any longer. One observes that the allowed channel for the
exact binding energy is now considerably smaller.

\begin{figure}
\epsfxsize=14cm \epsfbox{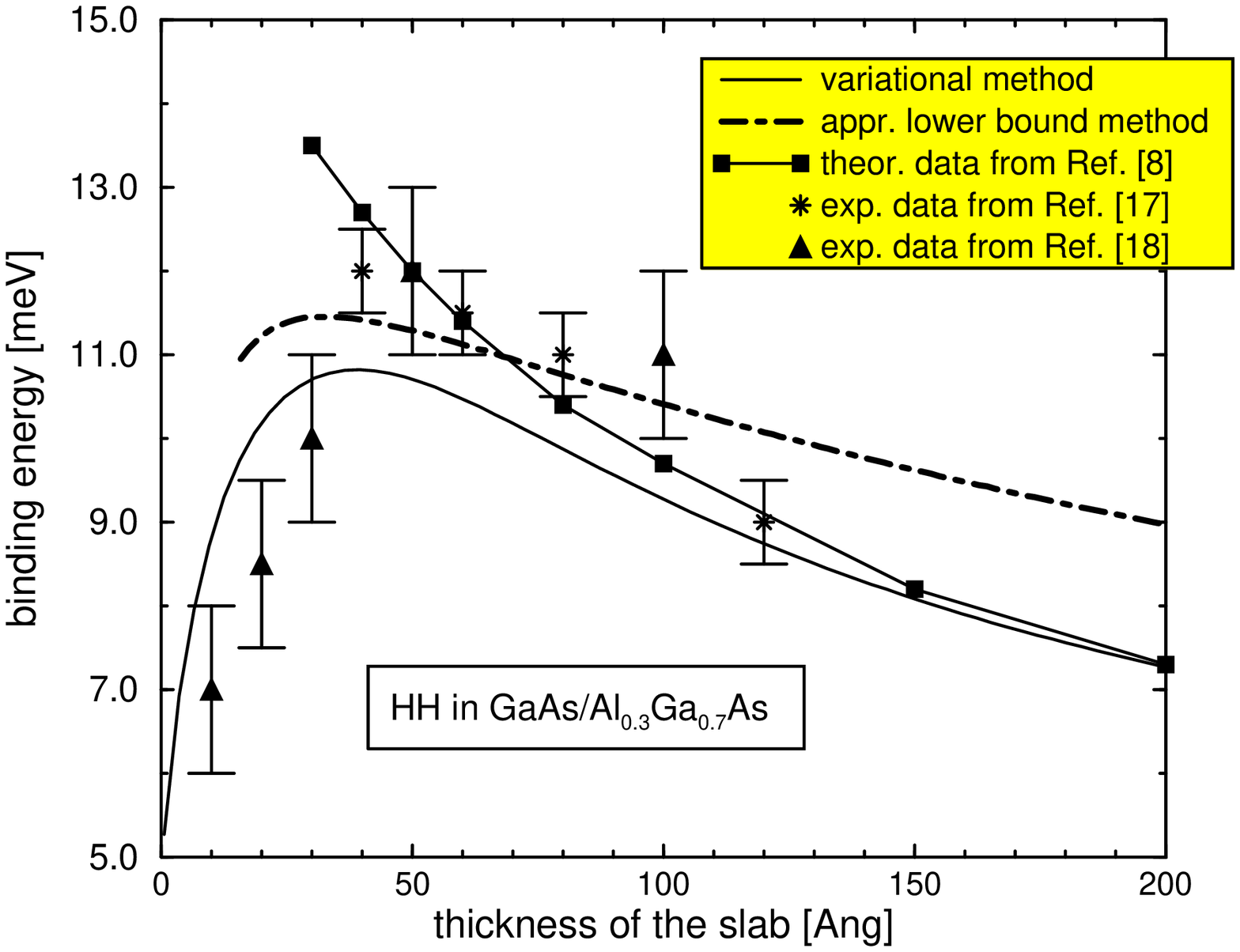}
\caption{Comparison of the present calculations (solid and
dashed-dotted lines) with experimental data (stars and triangles) as
well as with theoretical data (square symbols) for the
heavy-hole-exciton in $Al_{0.3}Ga_{0.7}As/GaAs/Al_{0.3}Ga_{0.7}As$.}
\label{exc6}
\end{figure}

\begin{figure}
\epsfxsize=14cm \epsfbox{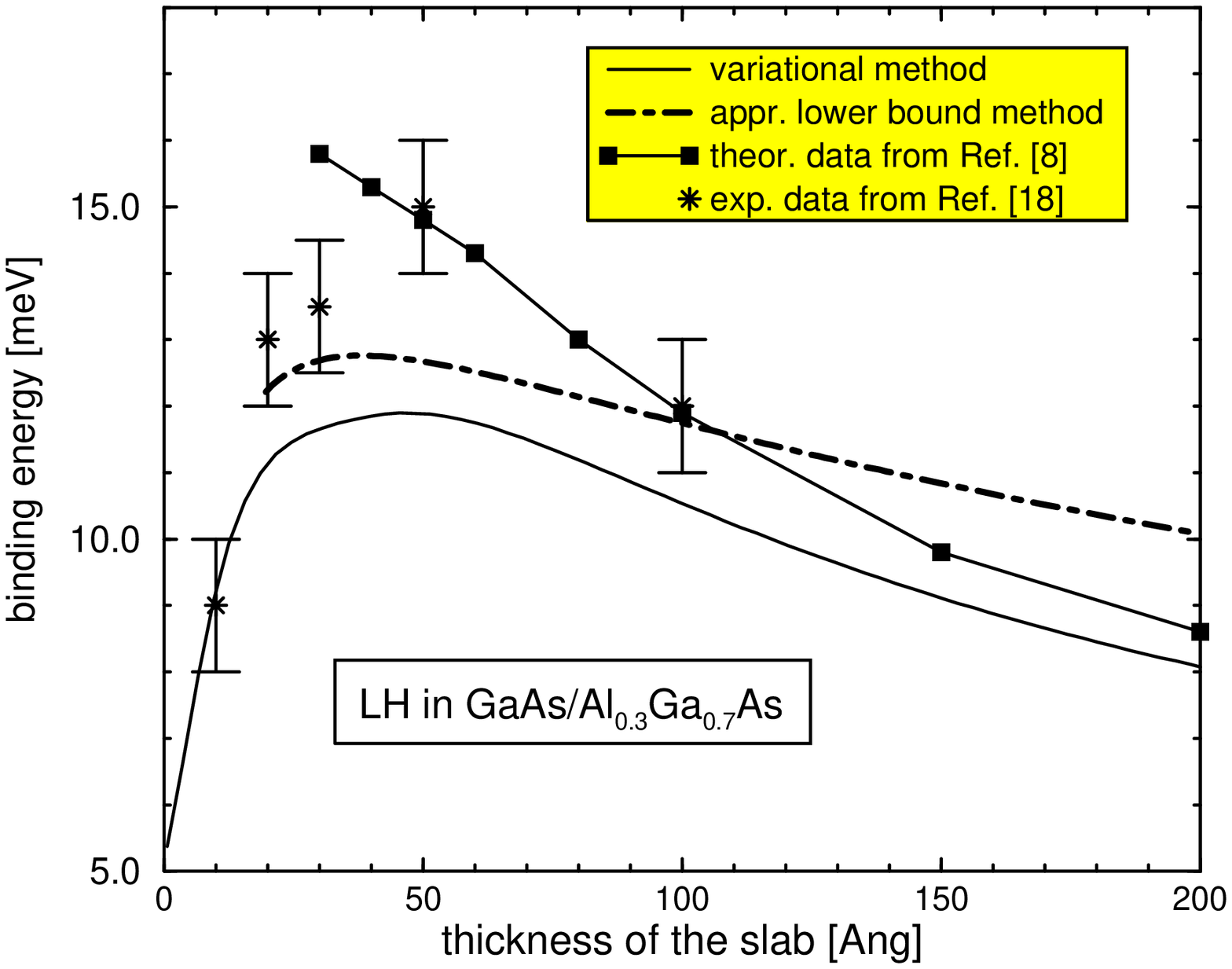}
\caption{Comparison of the present calculations (solid and
dashed-dotted lines) with experimental data (stars) as well as with
theoretical data (square symbols) for the light-hole-exciton in
$Al_{0.3}Ga_{0.7}As/GaAs/Al_{0.3}Ga_{0.7}As$. }
\label{exc7}
\end{figure}

Finally, Figs.~\ref{exc6},~\ref{exc7} compare experimental and
theoretical results, again for
$Ga_{0.7}Al_{0.3}As/GaAs/Ga_{0.7}Al_{0.3}As$. The observed peak
structure is reproduced by the present theory, but the experimental
binding energies are up to 1 meV larger. There are indications that
this discrepancy may be caused by our simplification of the real band
structure.  An enhancement of the exciton binding energies due to
valence band degeneracy and conduction band nonparabolicity has been
found by several authors (see, e.g., Refs.
\onlinecite{Andreani,Bandstructure}).

\section{conclusions}
The intention of this work was to calculate the binding energy of an
exciton in a rectangular quantum well with a higher reliability as was
previously to be found in the literature. In doing so, we provided
upper and lower bounds for the binding energy, which have a similar
shape and constitute an allowed channel for the exact binding energy
which is satisfactorily small. We can conclude that the peaks of the
binding energy as function of the well width are well established
within our model, and for a wide parameter range. In the case of
$Ga_{1-x}Al_{x}As/GaAs/Ga_{1-x}Al_{x}As$, our results revise those of
Ref.~\onlinecite{Andreani} thereby extending the regime of
``admissable" well widths to less than $30\, \AA$. From a
methodological point of view, our derivation of upper bounds on the
binding energy (or, equivalently, lower bounds to the ground-state
energy) can be transferred to related problems.

\acknowledgments
The authors would like to thank E. Bratkovskaya and H. Leschke
for inspiring discussions at earlier stages of the present study.
Furthermore, we are indebted to G. Flinn for critical remarks.
Financial support of the Heisenberg-Landau program (JINR-Germany
collaboration in theoretical physics) as well as Deutsche
Forschungsgemeinschaft (Graduiertenkolleg GRK 50/2) is gratefully
acknowledged.

\appendix
\section*{Upper and lower bounds for the anisotropic Coulomb problem}

To evaluate the expression (\ref{bound2}) for the lower bound on the
binding energy of Hamiltonian (\ref{h2}), we had to provide an
expression for the ground-state energy $e(A)$ of the Hamiltonian

\begin{eqnarray}
h_A &:=& -\vec \nabla^{\,2} - {2\over \sqrt{r_{\perp}^2 + Az^2}}\nonumber \\
&=& -\vec \nabla^{\,2} -
{2\over \sqrt{Ar^2 +(1-A)r_{\perp}^2}}.
\label{eq_a1}\end{eqnarray}

\noindent
We shall now briefly comment on this problem. Clearly, $e(A)$
interpolates between the two- and three-dimensional limits of the
hydrogen system. If $A$ is changed from $1$ to $0$, the ground-state
energy $E_A$ varies from -1 to -4 (in Rydberg units). Therefore, it is
tempting to choose a variational wave function for the ground state as
follows (see Ref.~\onlinecite{Gerpol}):

\begin{eqnarray}
\psi = C e^{-\lambda (r+ar_{\perp})},
\label{eq_a2}\end{eqnarray}

\noindent
where $a,\lambda$ are variational parameters. This wave function is
asymptotically exact in the three- ($A\to 1, \lambda \to 1, a\to 0$)
and two-dimensional ($A\to 0, \lambda a\to 2, \lambda \to 0$) limits.
Calculating the expectation value of Hamiltonian (\ref{eq_a1}), one
finds $e(A) \leq e_{GP}(A)$, where

\begin{eqnarray}
e_{GP}(A) = \min_{a}(a^2-1) {V^2(A,a) \over V^2(1,a) - a^2 V^2(0,a)/4},
\label{eq_a3}\end{eqnarray}

\noindent
and

\begin{eqnarray}
V(x,a) = \int\limits^{\pi/2}_0 d\theta {\sin\theta \over
(1+a \sin\theta)^2 \sqrt{\sin^2\theta + x \cos^2\theta}}.
\label{eq_a4}\end{eqnarray}

For a derivation of these equations and a discussion of the quality
of the bound $e_{GP}(A)$, we again refer to Ref.~\onlinecite{Gerpol}.
In Sec.~IV we made repeated use of formula (\ref{eq_a3}).

Interestingly enough, the same trial wave function provides us with a
lower bound for $e(A)$. To demonstrate this, we start from the identity

\begin{eqnarray}
-\vec \nabla^{\,2} \psi + V_{tr}\psi &=& - \lambda^2 (1+a)^2\,\psi, \nonumber \\
V_{tr} &=& -\left[{2\lambda \over r} + {a\lambda \over r_{\perp}}+
2a\lambda^2 \left(1-{r_{\perp} \over r}\right)\right].
\label{eq_a5}\end{eqnarray}

Thus, $\psi$ is an eigenfunction of the Hamiltonian $h_{tr} = -\vec
\nabla^{\,2} +V_{tr}$, corresponding to the eigenvalue $e_{tr} = -
\lambda^2 (1+a)^2$. In fact, $\psi$ is the ground-state eigenfunction
of $h_{tr}$. To prove this, one should realize that i) the ground-state
of $h_{tr}$ is nondegenerate for all $a$, and ii) $e_{tr}$ coincides
with the ground-state energy for $a=0$; besides, the chosen $\psi$ has
no zeros. We remark that $h_{tr}$ has the same two- and
three-dimensional limits as $h_A$.

Now let $\psi_A$ be the exact ground-state wave function of the
Hamiltonian $h_A$. Then we obtain

\begin{eqnarray}
e(A) &=& \langle \psi_A | h_A|\psi_A \rangle\nonumber\\
&=&\langle \psi_A | h_{tr} - {2\over \sqrt{Ar^2 +(1-A)r_{\perp}^2}} -
V_{tr}|\psi_A\rangle  \nonumber \\
& \geq & e_{tr} + \langle \psi_A |
{2\lambda \over r} + {a\lambda \over r_{\perp}}+
2a\lambda^2 \left(1-{r_{\perp} \over r}\right)
- {2\over \sqrt{Ar^2 +(1-A)r_{\perp}^2}} |
\psi_A\rangle \nonumber \\
& \geq & e_{tr} + \langle \psi_A |{1\over r}\left[
2\lambda + {a\lambda \over \rho}
- {2\over \sqrt{A +(1-A)\rho^2 }}\right] |\psi_A\rangle,
\label{eq_a6}
\end{eqnarray}

\noindent
where we defined $\rho := r_{\perp}/r$ and made use of $\rho \leq 1$.
The right-hand side of the latter inequality (\ref{eq_a6}) may be
considered as a function of the variational parameters $\lambda$ and
$a\lambda$ ($A$ being the experimental ``input"). We evaluate this
function as follows. Firstly, we calculate the minimum of the
expression in square brackets as a function of $\rho$; inserting the
corresponding solution into inequality (\ref{eq_a6}), the square
bracket is a pure $c$-number and can be extracted from the expectation
value. Secondly, we choose $\lambda$ such that the extracted square
bracket vanishes and, finally, $a\lambda$ such that the lower bound
assumes a maximum. This leads us to

\begin{equation} \label{eq_a7}
e(A) \geq -{4\over 1+3A}.
\end{equation}

Obviously, this lower bound gives the correct values $e(1)=-1$
and $e(0) = -4$.

\begin{figure}
\epsfxsize=14cm \epsfbox{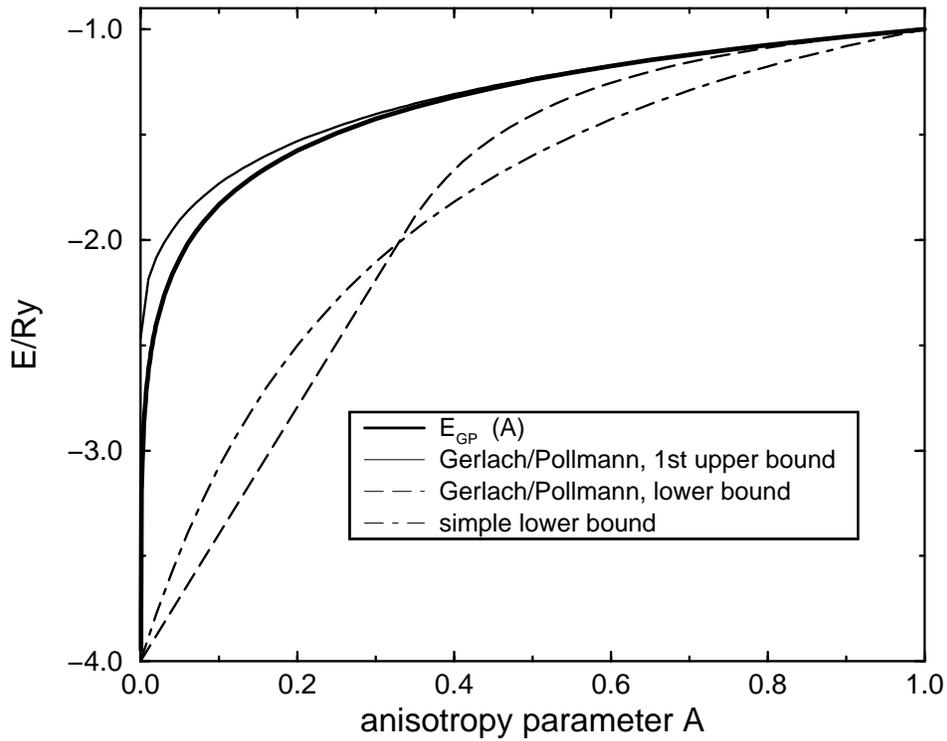}
\caption{Upper (solid lines) and lower (dashed lines) bounds for the
anisotropic Coulomb problem. The solution which is considered as
numerically exact is drawn as a thick line.}
\label{anis} \end{figure}

\end{document}